\documentstyle[twocolumn]{jpsj} 
\input epsf.tex
\def\lsim{\mathop {\vtop {\ialign {##\crcr 
$\hfil \displaystyle {<}\hfil $\crcr \noalign {\kern1pt \nointerlineskip } 
$\,\,\sim$ \crcr \noalign {\kern1pt}}}}\limits}
\def\gsim{\mathop {\vtop {\ialign {##\crcr 
$\hfil \displaystyle {>}\hfil $\crcr \noalign {\kern1pt \nointerlineskip } 
$\,\sim$ \crcr \noalign {\kern1pt}}}}\limits}

\def\runtitle{
Coupled CDW and SDW Fluctuations as an Origin of Anomalous Properties 
of Ferromagnetic Superconductor UGe$_{2}$
}
\def\runauthor{Shinji {\sc Watanabe} and Kazumasa {\sc Miyake}
}

\title
{Coupled CDW and SDW Fluctuations as an Origin of Anomalous Properties 
of Ferromagnetic Superconductor UGe$_{2}$
}

\author
{
Shinji {\sc Watanabe}\footnote{
Present address: 
Institute for Solid State Physics, 
University of Tokyo, 5-1-5, Kashiwanoha, Kashiwa, Chiba 277-8581; 
e-mail:~swata@issp.u-tokyo.ac.jp}
and Kazumasa {\sc Miyake}
}

\inst
{
Department of Physical Science, 
Osaka University, Toyonaka, 560-8531, Japan
} 

\recdate
{
\today
}

\abst
{ 
It is shown that anomalous properties of UGe$_2$ can be understood 
in a unified way on the basis of a single assumption that the 
superconductivity is mediated by the coupled SDW and CDW fluctuations 
induced by the imperfect nesting of the Fermi surface 
with majority spins at $T=T_{\rm x}(P)$ deep in the ferromagnetic 
phase.  
Excess growth of uniform magnetization is shown to develop in the temperature 
range $T<T_{\rm x}(P)$ 
as a mode-coupling effect of coupled growth of 
SDW and CDW orderings, which has been observed by 
two different types of experiments. 
The coupled CDW and SDW fluctuations are shown to be essentially 
ferromagnetic spin fluctuations which induce a spin-triplet 
p-wave attraction. 
These fluctuations consist of two modes, spin and charge fluctuations 
with large momentum transfer of the nesting vector. 
An anomalous temperature dependence of the upper 
critical field $H_{{\rm c}2}(T)$ such as 
crossing of $H_{\rm c2}(T)$ at $P=11.4$ kbar and $P=13.5$ kbar, 
can be understood by the strong-coupling-superconductivity formalism. 
Temperature dependence of the 
lattice specific heat including a large shoulder near $T_{\rm x}$ is also 
explained quite well as an effect 
of a kind of Kohn anomaly associated with coupled CDW-SDW transition.  
}

\kword
{
ferromagnetic superconductor, 
coupled SDW and CDW fluctuations, UGe$_2$
}

\begin{document}
\sloppy
\maketitle

\section{INTRODUCTION}
\begin{figure}
\begin{center}
\epsfxsize=7cm \epsfbox{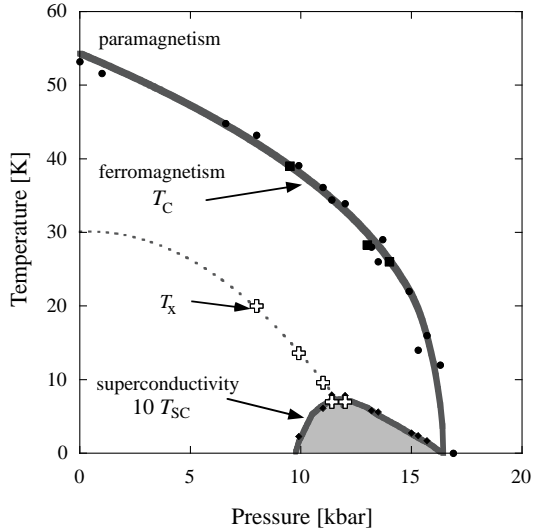} 
\end{center}
\caption{
Temperature-pressure phase diagram of 
UGe$_2$~\cite{huxley}.  
$T_{\rm C}$ and $T_{\rm SC}$ 
denote Curie temperature and superconducting transition 
temperature, respectively.  The cusp anomaly in the resistivity is observed at 
$T=T_{\rm x}$ (see text). 
}
\label{fig:TPphase} 
\end{figure}

\begin{figure}
\begin{center}
\epsfxsize=7cm \epsfbox{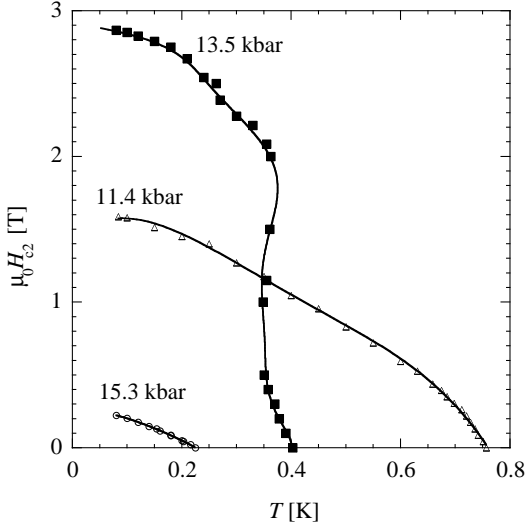}
\end{center}
\caption{
Upper critical field for the field applied in parallel to the 
easy axis ($a$-axis) at various pressures in UGe$_2$~\cite{huxley,sheikin}.  
}
\label{fig:Hc2} 
\end{figure}

The recent finding of the ferromagnetic superconductor 
UGe$_{2}$ has been attracting a great interest as the first material 
where both the superconducting (SC) state and the ferromagnetic (FM) 
state are sustained by $5f$ electrons~\cite{saxena}.  
Contrary to the first expectation, it soon turned out that the 
mechanism of the superconductivity has nothing to do with the 
ferromagnetic spin fluctuations associated with FM transition.  
Indeed, the phase diagram of UGe$_2$ presented in Fig.~\ref{fig:TPphase}, 
where $T_{\rm C}$ and $T_{\rm SC}$ denote the Curie temperature and 
the SC transition temperature, respectively, 
shows that the SC phase is located deep in 
the FM phase and a steep increase of $T_{\rm C}$ around $P=16.5$ kbar 
suggests the first order transition for the ferromagnetic 
transition~\cite{huxley}.  This is in marked contrast with the case of 
the usual {\it unconventional} SC phases which appear in close proximity to 
and around the magnetic phase boundary since the superconductivity is 
mediated by the enhanced magnetic fluctuations around the second-order 
phase transition.  
The upper critical field ($H_{{\rm c}2}(T)$) at $T\to 0$ 
far exceeds the so-called Clogston limit for singlet pairing, 
$H_{{\rm p}}(0)\simeq 1.84T_{\rm SC}$ (Tesla/K)
\cite{Clogston}, as seen 
in Fig.~\ref{fig:Hc2}.  So, it may be natural to confine ourselves within 
a triplet manifold.  
While phenomenological theories based on the ferromagnetism 
and superconductivity appeared recently\cite{machida,fomin}, 
the mechanism of superconductivity 
consistent with other fundamental properties has not been clarified yet.  

The superconducting mechanism appears to be related to another 
``phase boundary" denoted by $T_{\rm x}(P)$ in Fig.~\ref{fig:TPphase} 
where a slight decrease of the resistivity has been 
observed~\cite{saxena,huxley,tateiwa}. 
As pressure increases, $T_{\rm x}$ goes down similarly to $T_{\rm C}$ 
and merges into the vicinity of the maximum $T_{\rm SC}$ 
around $P=11.4$ kbar.  So, 
it is natural to consider that the fluctuations related to $T_{\rm x}$ 
plays an essential role to cause the superconductivity in UGe$_2$ as 
in the other {\it unconventional} superconductivity.  
Although the phase boundary was first identified by the small anomaly of 
the temperature dependence of the resistivity $\rho(T)$, 
it has begun to be recognized quite recently 
and turned out by new measurement that the uniform magnetization grows 
extra (other than the well developed ferromagnetic moment) 
for $T<T_{\rm x}$\cite{huxley,huxley2,tateiwa2,sato}.  
Such metamagnetic behaviors are observed even for $P>P_{\rm x}$, $P_{\rm x}$ 
being defined by $T_{\rm x}(P_{\rm x})=0$, under the magnetic field.  Indeed, 
$T_{\rm x}(H)$ reappears under the magnetic field and increases as a rate 
of about 1 K/T near the critical pressure 
$P=P_{\rm x}$~\cite{tateiwa2,sheikinprivate}.  

Furthermore, remarkable anomalies which characterize the superconductivity 
of UGe$_2$ have been observed in $H_{{\rm c}2}$ 
measurement: 
As shown in Fig.~\ref{fig:Hc2}, 
$H_{{\rm c}2}$ exhibits quite unusual $T$-dependence~\cite{huxley,sheikin}.  
Namely, $H_{{\rm c}2}$ at $P=11.4$ kbar for which $T_{\rm SC}$ takes maximum 
merely shows gradual increase, while 
$H_{{\rm c}2}$ at $P=13.5$ kbar shows first-order like steep increase 
although $T_{\rm SC}$ at $P=13.5$ kbar is smaller than that at $P=11.4$ kbar. 
This behavior can be also interpreted as the occurrence of 
the suppression of $H_{{\rm c}2}(T\to 0)$ at $P=11.4$ kbar 
by the magnetic field.  Such unusual aspects can be 
understood if the superconductivity is induced by the fluctuations 
associated with the transition at $T_{\rm x}$ as will be discussed 
in detail in this paper.  

So far, the nature of the transition at the phase boundary 
$T=T_{\rm x}(P)$ has not 
been established yet.  However, there are some circumstantial evidences 
that the transition may be the coupled SDW and CDW onset due to 
the imperfect nesting of the majority-spin band.  

The first evidence is that 
the Fermi surfaces of the majority-spin band obtained by the band-structure 
calculations is nearly nested both in the $a$- and 
$c$-direction~\cite{yamagami,pickett}.  
Although the first stage of calculation has been done on the basis of 
incorrect crystal symmetry due to the confusion of experiments, 
recent calculation by H. Yamagami, taking the correct crystal symmetry and 
correct easy-axis ($a$ axis) of magnetization, also shows that 
the Fermi surfaces are mainly composed of $5f$ electrons and 
have the structure which favors the imperfect nesting 
in the planes perpendicular to the $a$- and 
$c$-axis\cite{yamagami2}.  
So, it is expected that the nesting tendency is realized in some direction 
connecting the flat part of the Fermi surfaces leading to the enhanced 
CDW fluctuations of the majority-spin band around $T=T_{\rm x}$.  
It is noted that CDW fluctuations of majority-spin band 
is inevitably accompanied by SDW fluctuations.  
This picture is verified by the fact that the ferromagnetic moment of the 
unit cell is almost constant, $M\simeq$ 1.0$\mu_{\rm B}$, along 
$T_{\rm x}$~\cite{huxley,tateiwa2,sato}.  

The second one is the existence of large lattice entropy observed 
by the specific heat $C$ below $T\sim T_{\rm x}<T_{\rm C}$.  
Integrating $C/T$ with respect to $T$ from $T=0$ K to $T_{\rm C}$, 
the molar entropy is estimated to be 
22 J/mole$\simeq$2.6$\times N_{\rm A}k_{\rm B}$ which is far larger than 
the entropy associated with spin degrees of freedom, 
$\ln 2\times N_{\rm A}k_{\rm B}$.  
The molar entropy due to the Fermi quasiparticles 
is estimated to be 1.7 J/mole by means of the Sommerfeld constant 
$\lim_{T\to 0}C/T\simeq$ 0.0314 J/mole$\cdot$K.  This value is much smaller 
than the observed one.  
The entropy due to the Debye phonon also gives only 1/3 of the observed one  
(see the discussion in \S3).  
Then the extra entropy should be ascribed to the other degrees of freedom.  
One probable candidate is the softened optical modes coupled to 
the CDW instability of the majority-spin band.  

The third one is the fact that the lattice structure of UGe$_2$ is 
similar to that of $\alpha$-Uranium in which the CDW has been confirmed 
experimentally by neutron scattering in consistent with 
the nesting vector calculated by the band theory~\cite{mermeggi}. 
$\alpha$-Uranium has the zig-zag chain arrangement of U atoms along the c axis. 
UGe$_2$ also has the the zig-zag chain arrangement of U atoms along the a axis 
and Ge atoms are located between U atoms. 
Moreover, the $T$-dependences of $\rho(T)$ and $C(T)$ of $\alpha$-Uranium are 
similar to those of UGe$_2$: 
The cusp anomaly of $\rho(T)$ and the convex curvature of $C(T)$ at 
$T_{\rm CDW}$ 
have been also observed in $\alpha$-Uranium. 
Here, one may expect that $\rho(T)$ should raise if the CDW ordering occurs. 
Of course in the case of the perfect nesting, the gap is created at the entire 
region of the Fermi surface so that the system becomes insulator. 
However, in the case of the imperfect nesting, the situation is not so simple.  
The $T$-dependence of resistivity is determined by the competition 
between two opposite factors: 
One is the tendency that makes the system insulator by the gap creation at 
some parts 
of the Fermi surface. 
The other is the effect of the reduction of the scattering amplitude among 
quasi particles by the CDW ordering that makes the resistivity lower. 
We remark that $\rm NbSe_2$ where the CDW ordering due to the imperfect nesting 
has been confirmed experimentally also shows the similar $\rho(T)$ 
which decreases below $T_{\rm CDW}$~\cite{NbSe2}.  

The fourth one is a rather sharp decrease of the residual resistivity 
$\rho(T\to 0)$ across $P=P_{\rm x}$ when $P$ is 
increased~\cite{huxley,settai}, 
implying that the area of the Fermi surface recovers as the coupled CDW and 
SDW ordering vanishes at $P=P_{\rm x}$.  This is because the $\rho(T\to 0)$ 
depend directly on the carrier number or the area of the Fermi surface and the 
scattering mechanism is essentially unchanged across $P=P_{\rm x}$,  
in contrast to the case where $T$ changes across $T_{\rm x}$ with the pressure $P$ fixed.  

In this paper, we show that fundamental properties of UGe$_2$ mentioned above 
can be understood in a unified picture that 
the superconductivity is mediated by coupled CDW and SDW fluctuations 
due to imperfect nesting in the majority-spin band at $T=T_{\rm x}$. 
In \S2, we show that the growth of extra magnetization 
for $T<T_{\rm x}$ is 
understood as a mode-coupling effect of the coupled CDW-SDW ordering.  
In \S3, we show that the anomalously large entropy around 
$T=T_{\rm x}(P=0)$ 
can be explained as contributions from optical phonons softened 
by the Kohn effect due to the onset of CDW ordering of the majority-spin band.  
In \S4, we show that the pairing interaction among quasiparticles 
in the majority-spin band is induced by the {\it ferromagnetic} 
fluctuations associated with coupled CDW and SDW fluctuations 
enhanced around $P=P_{\rm x}$.  
Namely, the spin susceptibility $\chi_{\rm s}(q\sim 0,{\rm i}\omega_m)$ is 
enhanced by the mode-coupling effect of CDW and SDW fluctuations 
just as these give rise to the excess uniform magnetization below 
$T<T_{\rm x}$.  These fluctuations induce the spin-triplet ``p"-wave 
paring.  It is shown that the anomalous temperature dependence 
of the upper critical field $H_{{\rm c}2}(T)$ observed at 
$P=11.4$ kbar and $P=13.5$ kbar 
can be understood by the strong-coupling formalism on the basis of 
the present mechanism and the experimental 
fact that $T_{\rm x}$ is an increasing function of $H$~\cite{sheikin}.  
Throughout this paper we use the unit such that the 
Boltzmann constant $k_{\rm B}=1$ and the Plank constant $\hbar=1$ 
unless they are explicitly written down.

\section{Mode-coupling between CDW and SDW}
In this section, we show that the excess growth of uniform magnetization 
can occur below $T<T_{\rm x}$ if the transition at $T=T_{\rm x}$ is the 
coupled CDW-SDW ordering.  Such a growth of the magnetization in UGe$_2$ 
has been observed by elastic neutron scattering\cite{huxley}, and 
direct measurements of magnetization by ac-method\cite{huxley,tateiwa2,sato}.  
In other words, it turns out that a central assumption of the present 
scenario is supported by a direct experimental evidence.  

\begin{figure}
\begin{center}
\epsfxsize=8cm \epsfbox{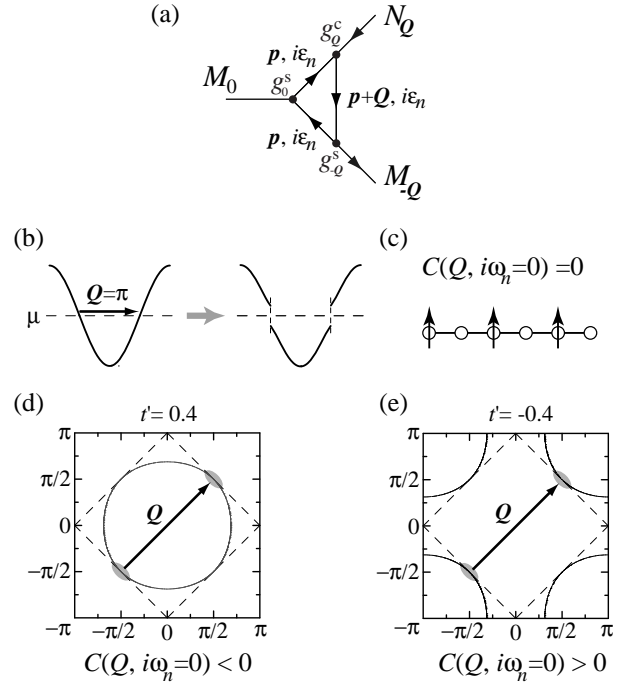} 
\end{center}
\caption{
(a) The mode-coupling term between the CDW and the SDW 
in the lowest order. 
The band picture (b) and the real-space picture (c)
by the perfect nesting in the majority-spin band 
in the one-dimensional simple lattice. 
The Fermi surface with $\mu=0$ (solid line) and 
the nesting zone boundary (dashed line) 
in the two-dimensional square lattice 
in the majority-spin band
$
\varepsilon({\bf p})=-2 \{ \cos(p_{x})+\cos(p_{y}) \}
-4t' \cos(p_{x})\cos(p_{y})
$
with (d) $t'=0.4$ and (e) $t'=-0.4$. 
The imperfect nesting by vector ${\bf Q}=(\pi,\pi)$ 
causes the gap at the shaded spots.
}
\label{fig:modeCS}
\end{figure}

An ingredient of such a growth of magnetization below $T<T_{\rm x}$ is 
the fact that in the free energy there appears a mode-coupling term among 
the uniform magnetization, 
$M_{0}$, the order parameter of CDW, $N_{\bf Q}$, and of SDW, 
$M_{\bf -Q}$, with a nesting vector {\bf Q}, in the region 
$T<T_{\rm x}$: 
\begin{eqnarray}
F=F_{0}+[C({\bf Q},i\omega_{n}=0)M_{0}N_{\bf Q}M_{\bf -Q}+ {\rm c.c.}],
\label{eq:free-energy}
\end{eqnarray}
where $F_{0}$ denotes the free energy other than the mode-coupling term.  
%
Here we note that the mode-coupling term with uniform charge $N_{0}$, 
$M_{\bf Q}$ and $M_{\bf -Q}$ can also appear in the free-energy expansion. 
However, this term is considered to be suppressed by the 
long-range Coulomb interaction which works to maintain the charge 
neutrality. 
%
The mode-coupling coefficient $C({\bf Q},i\omega_{n}=0)$ 
in eq.~(\ref{eq:free-energy}) can be 
calculated in terms of the Feynman diagram shown in Fig.~\ref{fig:modeCS}(a):
\begin{eqnarray}
& &C({\bf Q},i\omega_{n}=0)=g^{\rm s}_{0}g^{\rm c}_{\bf Q}g^{\rm s}_{-\bf Q}
\nonumber \\
&\times& 
\int \frac{d^{3}p}{(2\pi)^3}
\frac{1}{4T}\cosh^{-2}\bigl(\frac{\varepsilon({\bf p})-\mu}{2T}\bigr)
\frac{1}{\varepsilon({\bf p})-\varepsilon({\bf p}+{\bf Q})}, 
\label{eq:modeCoef}
\end{eqnarray}
where $\varepsilon({\bf p})$ is the dispersion of 
the majority-spin band and 
$\mu$ is the chemical potential.  
The coefficients in eq.~(\ref{eq:modeCoef}), $g^{\rm s}_{0}(>0)$, 
$g^{\rm c}_{\bf Q}(>0)$, and $g^{\rm s}_{-{\bf Q}}(>0)$, denote the coupling 
between quasiparticles and spin fluctuations with $q=0$, charge fluctuations 
with ${\bf q}={\bf Q}$, 
and spin fluctuations with ${\bf q}=-{\bf Q}$, respectively.  

First, we consider the perfect-nesting case. 
The perfect nesting occurs in the bipartite lattices 
at half filling with $\mu=0$. 
In this case, the relation 
$\varepsilon({\bf p}+{\bf Q})=-\varepsilon({\bf p})$
holds and the mode-coupling coefficient 
is given as follows: 
%
\begin{eqnarray}
C({\bf Q},i\omega_{n}=0)& &=g^{\rm s}_{0}g^{\rm c}_{\bf Q}g^{\rm s}_{-{\bf Q}}
\Omega\int d\xi D(\xi)
\frac{1}{2\xi}
\frac{1}{4T}\cosh^{-2}\bigl(\frac{\xi}{2T}\bigr)
\nonumber
\\
& &\simeq
g^{\rm s}_{0}g^{\rm c}_{\bf Q}g^{\rm s}_{-{\bf Q}}\Omega
\frac{dD(\xi)}{d\xi}\bigl|_{\xi=0}
\label{eq:coef}
\end{eqnarray}
where 
$D(\xi)\Omega=\int d^{3}p/(2\pi)^{3}
\delta(\varepsilon({\bf p})-\mu-\xi)$ 
is the density of states measured from at the Fermi level, $\mu$, 
with $\Omega$ being the volume of the system.  
The last approximate equality 
in eq.~(\ref{eq:coef})
is valid when $T_{\rm x}$ is assumed to be 
much smaller than the bandwidth of quasiparticles.  
Therefore, the coefficient $C$ 
vanishes as far as the $D(\xi)$ is symmetric function of $\xi$ 
at $\xi\sim 0$.  This result can be understood as follows:  
As is well known, the number density of the majority 
spin band, under the fixed chemical potential, remains unchanged even under 
the growth of the coupled CDW and SDW ordering, 
$\Delta_{\rm C-S} \propto M_{0}N_{\bf Q}M_{\bf -Q}$, in such a case.  
This can easily be seen 
if one considers the one-dimensional simple lattice 
as an example in Fig.~\ref{fig:modeCS}(b): 
By the CDW-SDW transition 
the rearrangement of the charges with majority spins occurs
but the net magnetization does not change 
as shown in Fig.~\ref{fig:modeCS}(c). 

In case that $D(\xi)$ has a weak asymmetric part with respect to $\xi=0$, 
the coefficient $C$ does not vanish 
so that the number density of majority spin band changes in general.  
However, it may be reasonable to assume $D(\xi)$ is symmetric around 
$\xi=0$ in the case where the perfect nesting is realized as in 
two-dimensional(2D) and three dimensional(3D) Hubbard model 
with only nearest neighbor hopping.  In one-dimensional case, 
however, $D(\xi)$ has a weak asymmetry in general under the condition of 
perfect nesting.  In such a case, the sign of the magnetization change 
cannot be determined by the sign of the term of 
$\cal{O}$$(\Delta_{\rm C-S}^{2})$, but is determined by the sign of 
the term with logarithmic correction, $\Delta_{\rm C-S}^{2}
\ln\Delta_{\rm C-S}$.  In any case, the effect is 
small one proportional to $|dD(\xi)/d\xi|_{\xi=0}|\ll D(\xi=0)$, in contrast 
to the case of inperfect nesting where the effect is proportional to 
$D({\xi=0})$ as will be discussed below.  

Next we consider the imperfect-nesting case.
In this case, the energy gap opens at the Fermi surface 
on the zone boundary 
transferred 
by the nesting vector $\bf Q$.
Hence, the coefficient eq.~(\ref{eq:modeCoef}) does not diverge 
but has a finite value. 
The important point is that the sign of 
$C({\bf Q},i\omega_{n}=0)$ changes according to the relative 
positions between the Fermi surface and the nesting zone boundary.

As an example let us consider 
the majority-spin band on the two-dimensional square lattice 
\begin{eqnarray}
\varepsilon({\bf p})=-2t \{ \cos(p_{x})+\cos(p_{y}) \}
-4t' \cos(p_{x})\cos(p_{y}),
\nonumber
\end{eqnarray}
with the nearest-neighbor hopping $t$ and 
the next-nearest-neighbor hopping $t'$. 
In Fig.~\ref{fig:modeCS}(d) and Fig.~\ref{fig:modeCS}(e) 
the Fermi surface with $\mu=0$ and 
nesting zone boundary is plotted by the solid line and the dashed 
line, respectively. 

For positive $t'$ the Fermi surface with $\mu=0$ is located  
inside the nesting zone boundary as in Fig.~\ref{fig:modeCS}(d). 
In this case the mode-coupling coefficient has a negative value 
$C({\bf Q},i\omega_{n}=0)<0$. 
Namely, in this case 
the uniform magnetization $M_{0}$ increases for $T<T_{\rm x}$.  
This can be understood as follows: 
The energy gap opens at the part of the Fermi surface 
along the nesting zone boundary, which is drawn by 
the shaded spots in Fig.~\ref{fig:modeCS}(d).  By the gap creation, 
the energy band inside the zone boundary shifts downward. 
As for the area at which the gap opens the situation is the same as 
the perfect nesting case. 
Namely, the shaded spots in Fig.~\ref{fig:modeCS}(d) 
do not contribute to the change of uniform magnetization. 
However, the important difference from the perfect-nesting case
is that the Fermi surface which deviates 
from the nesting zone boundary (i.e., the part where the gap does not open) 
shifts downward by the gap creation. 
Hence, 
the electrons flow into the majority-spin band from the minority-spin band 
so that the uniform magnetization increases.  

On the contrary, for negative $t'$ the Fermi surface with $\mu=0$ is 
located outside the nesting zone boundary as in Fig.~\ref{fig:modeCS}(e).  
In this case the energy band shifts upward by the gap creation 
and it is shown that the mode-coupling coefficient has a positive value 
$C({\bf Q},i\omega_{n}=0)>0$.  
Namely, in this case 
electrons flow into the minority-spin band from the majority-spin band 
so that the uniform magnetization $M_{0}$ decreases for $T<T_{\rm x}$.  

%
Here we have discussed the change of the uniform magnetization 
referring to the one- and two-dimensional lattices 
as typical examples to draw a general conclusion. 
In the real systems, the shape of the Fermi surface is often 
more complicated, but our statement derived here can be 
applied to the Fermi surface transferred by the nesting vector $\bf Q$.  
%
As for $\rm UGe_2$, the Fermi surface of the majority-spin band 
is located inside the nesting zone boundary according to the 
recent band-structure calculation~\cite{yamagami}. 
Namely, it seems to correspond to the case (d) in above examples, 
which should exhibit the increase of the uniform magnetization 
for $T<T_{\rm x}$ as observed experimentally
~\cite{huxley,tateiwa2,sato}.  

Concluding this chapter, let us remark that the mechanism of magnetization 
changes due to simultaneous ordering of CDW and SDW with the equal nesting 
wave vector 
is quite general one which may be found in other systems with such a 
property.  
It is also remarked that the sign of magnetization change depends on 
the curvature of the Fermi surface near the nesting vector as discussed 
above.  

\section{Specific heat due to Kohn anomaly}
\begin{figure}
\begin{center}
\epsfxsize=8cm \epsfbox{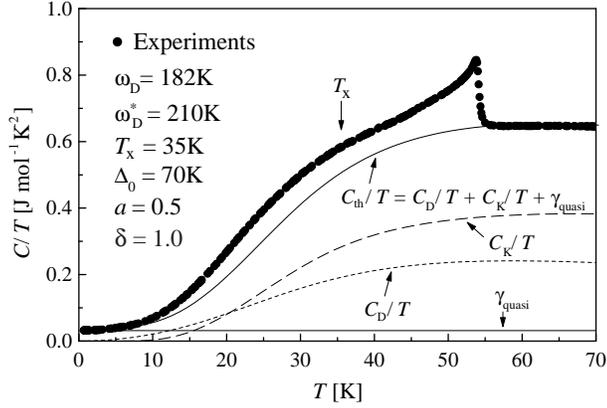} 
\end{center}
\caption{$C/T$ vs $T$.  The solid circles represent the experimental 
values\cite{huxley}, 
the solid line the theoretical ones, which consist of the Debye 
contribution (dotted line), the cotributions from optical modes with the 
Kohn anomaly (dashed line), and the fermionic contributions from 
quasiparticles (solid line).  
}
\label{fig:Gamma}
\end{figure}

In this section, we discuss how the temperature dependence of the specific 
heat $C(T)$ at $T<T_{\rm C}$, especially a gentle hump around $T=T_{\rm x}$, 
is understood from the present point of view that $T_{\rm x}$ is 
the onset temperature of coupled CDW-SDW ordering.  The specific heat 
due to the Debye phonon cannot afford to explain the global behavior 
of $C(T)/T$ as can be seen in Fig.~\ref{fig:Gamma} in which the Debye 
contribution 
$C_{\rm D}/T$ is compared with the experiment at ambient pressure 
\cite{huxley}.  
The Debye contribution is given as 
\begin{equation}
{C_{\rm D}\over T}=\Lambda\biggl[{4\over T^{2}}\int_{0}^{\omega_{\rm D}}{\rm d}
\omega{\omega^{3}\over e^{\omega/T}-1}-{\omega_{\rm D}^{4}\over T^{2}}
{1\over e^{\omega_{\rm D}/T}-1}
\biggr],
\label{debye1}
\end{equation}
with the coefficient $\Lambda$ defined by 
\begin{equation}
\Lambda\equiv{9\over2}{Nk_{\rm B}\over\omega_{\rm D}^{3}},
\label{debye2}
\end{equation}
where $N$ and $\omega_{\rm D}$ denote the unit-cell number included in the 
system and the averaged Debye energy, respectively.  
In the low temperature limit, $T\ll\omega_{\rm D}$, 
eq.~(\ref{debye1}) is reduced to the conventional form as 
\begin{equation}
{C_{\rm D}\over T}\simeq Nk_{\rm B}{12\pi^{4}\over5\omega_{\rm D}^{3}}T^{2}.  
\label{debye3}
\end{equation}
Comparing the experimental value for the coefficient of $T^{2}$ term of 
eq.~(\ref{debye3}), 
3.2$\times$10$^{-4}$ J/K$^{4}$$\cdot$mole \cite{huxley}, 
we obtain the Debye energy as 
$\omega_{\rm D}=182$~K.  Using this value for $\omega_{\rm D}$, we have  
ploted $C_{\rm D}/T$ in Fig.~\ref{fig:Gamma}.  
In order to account for the hump around $T_{\rm x}$, 
there should be an extra contribution to the entropy from 
other than the magnetic origin because the difference between 
$C_{\rm D}/T$ and the experimental value is far larger than the jump 
of $C/T$ at $T=T_{\rm C}$, the ferromagnetic transition.  

The optical phonons are subject to the Kohn anomaly if the Fermi surface of 
quasipartices is nested as expected in the majority-spin band of UGe$_2$.  
We parameterize the dispersion of optical phonons exhibiting the Kohn anomaly 
as follows:
\begin{equation}
\varepsilon({\bf k})=\sqrt{(s_{\parallel}k_z)^{2}+
s_{\perp}^{2}(k_x^{2}+k_y^{2})^{2}+\Delta^{2}}, 
\label{debye4}
\end{equation}
where the wave vector ${\bf k}$ is measured from the nesting vector 
${\bf Q}$, and 
we here take the direction along which the nesting occurs as 
$k_{z}$-direction without loss of generality.  
The gap $\Delta$ parameterizes the degree of softening 
of the phonons, and $s_{\parallel}$ and $s_{\perp}$ denote the sound 
velocities for longitudinal and transverse directions, respectively, 
with respect to ${\bf Q}$.  There exist two independent optical-phonon modes 
corresponding to two Ge ions in the unit cell.  For simplicity, we assume 
that the dispersions of two modes are identical.  
Then, the energy due to optical-phonon modes with Kohn anomaly is given as 
\begin{equation}
E_{\rm K}=\Xi\int_{0}^{\omega_{\rm D}^{*}}{\rm d}\omega_{z}
\int_{0}^{\omega_{\rm D}^{*}}{\rm d}\omega \omega
{\varepsilon(\omega_{z},\omega)
\over e^{\varepsilon(\omega_{z},\omega)/T}-1},
\label{debye5}
\end{equation}
where 
$\omega_{z}\equiv s_{\parallel}k_{z}$, 
$\omega\equiv s_{\perp}\sqrt{k_x^{2}+k_y^{2}}$, 
$\delta\equiv s_{\perp}/s_{\parallel}$.  The Debye energy 
$\omega_{\rm D}^{*}$ is different from $\omega_{\rm D}$ for the 
acoustic modes in general.  

\begin{figure}
\begin{center}
\epsfxsize=8cm \epsfbox{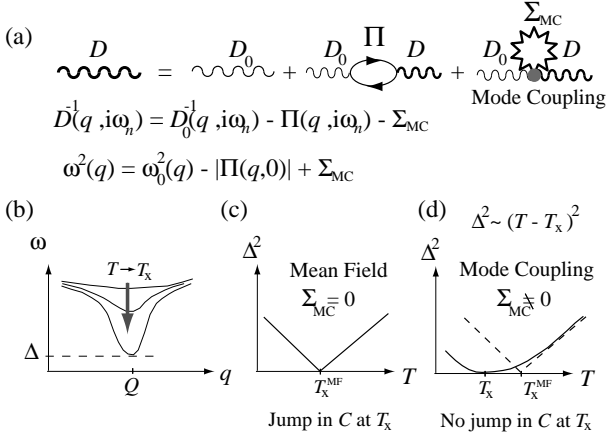} 
\end{center}
\caption{
(a) Dyson equation of phonon Green's function including the selfenergy of the 
mode coupling. 
(b) Softening of optical phonon mode at the nesting vector $\bf Q$. 
Temperature dependence of the gap in the mean-field framework (c) 
and in the framework including the mode-coupling term (d). 
}
\label{fig:spcfcMC}
\end{figure}

In the high temperature limit, $T\gg\omega_{\rm D}^{*}$, eq.~(\ref{debye5}) 
is reduced to 
\begin{equation}
E_{\rm K}=\Xi \int_{0}^{\omega_{\rm D}^{*}}{\rm d}\omega_{z}
\int_{0}^{\omega_{\rm D}^{*}}{\rm d}\omega \omega T
={1\over2}\Xi T\omega_{\rm D}^{*3}.
\label{debye6}
\end{equation}
The coefficient $\Xi$ is determined so as to reproduce 
the value of Dulong-Petit:
\begin{equation}
\Xi={6Nk_{\rm B}\over\omega_{\rm D}^{*3}}\times 2. 
\label{debye7}
\end{equation}
The specific heat $C_{\rm K}$ is obtained from eq.~(\ref{debye5}) as follows: 
\begin{eqnarray}
{C_{\rm K}\over T}
=\Xi \biggl[{1\over T^{3}}\int_{0}^{\omega^{*}_{\rm D}}{\rm d}\omega_{z}
\int_{0}^{\omega^{*}_{\rm D}}{\rm d}\omega \omega
{\varepsilon^{2}e^{\varepsilon/T}\over
(e^{\varepsilon/T}-1)^{2}}
\nonumber\\
\hspace*{-0.5cm}
-{1\over 2T}
{\partial\Delta^{2}\over\partial T}
\int_{0}^{\omega^{*}_{\rm D}}{\rm d}\omega_{z}
\int_{0}^{\omega^{*}_{\rm D}}{\rm d}\omega {\omega\over\varepsilon}
{(\varepsilon/T-1)e^{\varepsilon/T}+1\over
(e^{\varepsilon/T}-1)^{2}}\biggr].
\label{debye8}
\end{eqnarray}
The second term crucially depends on the temperature dependence of 
the gap $\Delta$ defined as the lowest optical-phonon energy 
in eq.~(\ref{debye4}).  
To discuss the relationship between the temperature dependence of the specific heat 
and the effect of the electron-phonon coupling, we show the Dyson equation for 
the phonon Green's function $D({\bf q}, {\rm i}\omega_{n})$ 
in Fig.~\ref{fig:spcfcMC}(a). 
Here 
$D_{0}({\bf q}, {\rm i}\omega_{n})$ denotes the bare Green's function for phonons. 
$\Pi$ and $\Sigma_{\rm MC}$ denote the electron polarization function and 
the selfenergy for the mode coupling, respectively. 
The softening of the optical-phonon mode with gap $\Delta$ 
at the wave vector $\bf Q$ which should be the same as the nesting vector 
is illustrated in Fig.~\ref{fig:spcfcMC}(b). 
On the mean-field level where 
$\Sigma_{\rm MC}$ is set to be zero, 
the $T$-dependence of $\Delta$ 
is given near the ordering temperature 
$T_{\rm x}^{\rm MF}$ like 
\begin{equation}
\Delta_{\rm MF}^{2}\simeq J|T-T_{\rm x}^{\rm MF}|, 
\label{gap1}
\end{equation}
where $J$ is a positive constant which is determined by the band structure of 
quasiparticles and the electron-phonon coupling constant.  
Then, $C_{\rm K}/T$ given by eq.~(\ref{debye8}) exhibits a discontinuous jump 
at $T=T_{\rm x}^{\rm MF}$ through the factor 
$\partial\Delta^{2}/\partial T$ as shown in Fig.~\ref{fig:spcfcMC}(c).  
However, in case the mode-coupling term cannot be neglected, 
in the limit $\Delta\to 0$ 
the mode-coupling effect gives 
the self-energy of the relevant optical phonon mode the term like 
$-M_{1}+M_{2}\Delta$, $M_1$ and $M_2$ being positive constants with weak 
temperature dependence near $T_{\rm x}$ as discussed in ref.~\cite{suzuki} 
(see Fig.~\ref{fig:spcfcMC}(d)).  
Thus the self-consistent equation 
for $\Delta$ takes the following form at $T>T_{\rm x}$, 
$T_{\rm x}$ being the ordering temperature determined self-consistently 
as below:
\begin{equation}
\Delta^{2}\simeq J(T-T_{\rm x}^{\rm MF})+M_{1}-M_{2}\Delta. 
\label{gap2}
\end{equation}
By solving this equation near 
$T_{\rm x}\equiv T_{\rm x}^{\rm MF}-(M_{1}/J)$, we obtain 
\begin{equation}
\Delta^{2}=J^{2}(T-T_{\rm x})^{2}/M_{2}^{2}. 
\label{gap3}
\end{equation}
For $T<T_{\rm x}$, similar temperature dependence is obtained by solving 
much more complicated equation due to the existence of ordering.  
Namely, it turns out that in case the effect of the mode coupling is relevant, 
the jump in the specific heat at the transition temperature $T_{\rm x}$ 
is smeared out by the fluctuations. 
We consider that UGe$_2$ may correspond to such a case 
where the mode-coupling term 
cannot be neglected for the fluctuations due to the strong electron-phonon coupling. 
Thus, to calculate the specific heat of UGe$_2$ 
we introduce a parameterization for $T$-dependence of  $\Delta$ 
as follows: 
\begin{equation}
\Delta={\Delta_{0}^{2}(T^{2}-T_{\rm x}^{2})^{2}\over
T_{\rm x}^{4}+aT^{4}}.
\label{debye9}
\end{equation}
This is reduced in low- and high-temperature limits as follows: 
\begin{equation}
\Delta^{2}(T)\simeq
\cases{
\Delta_{0}^{2}, &$T\ll T_{\rm x}$;\cr
\quad&\ \cr
\displaystyle{\strut{\Delta_{0}^{2}\over a}}, &$T\gg T_{\rm x}$.\cr}
\label{debye10}
\end{equation}
%

In Fig.~\ref{fig:Gamma}, the contribution from eq.~(\ref{debye8}) is shown as 
dashed line for parameters $\omega^{*}_{\rm D}$=210 K, 
$\Delta_{0}$=70 K, $a$=0.5, and $\delta$=1.0.  
One solid line in Fig.~\ref{fig:Gamma} is a contribution from 
the Fermi quasiparticles 
$\gamma_{\rm quasi}$ which is determined experimentally as 
$\gamma_{\rm quasi}=\lim_{T\to 0}C/T\simeq0.0314$ J/mole K$^2$~{\cite{huxley}.  
The other solid line is for $C_{\rm th}/T$ which is the sum of 
$C_{\rm D}/T$, $C_{\rm K}/T$, and $\gamma_{\rm quasi}$.  
The curve $C_{\rm th}/T$ reproduces quite well the global behavior of 
observed $C/T$.  
In Fig.~\ref{fig:Gamma2}, $C/T$ (solid circle), 
$C_{\rm th}/$ (solid line), and their difference 
$\Delta C/T\equiv C/T-C_{\rm th}/T$ (solid square) are shown.  
The origin of $\Delta C/T$ near $T=T_{\rm C}$ is apparently the 
magnetic entropy of local component of magnetization, while that of 
a hump around $T$=30 K may be attributed to the mass enhancement due 
to the criticality of coupled CDW-SDW ordering at $T=T_{\rm x}\simeq$35 K, 
where the enhancement of a few times of $\gamma_{\rm quasi}$ is expected 
as observed at $(T,P)\simeq (0,P_{\rm x})$\cite{tateiwa}.  

If we had not taken into account the effect of Kohn anomaly, we could not 
account for the gentle hump of $C/T$ around $T$=35 K by the Einstein model 
which simulates the two optical phonons with frequency $\omega_{\rm E}$.  
Indeed, its contribution $C_{\rm E}/T$ is given as 
\begin{equation}
{C_{\rm E}\over T}={6Nk_{\rm B}\over T^{3}}{\omega_{\rm E}
e^{\omega_{\rm E}/T}\over(e^{\omega_{\rm E}/T}-1)^{2}}.
\label{debye11}
\end{equation}
This has a rather sharp maximum 
$C_{\rm E}^{\rm max}/T\simeq 76/\omega_{\rm E}^{2}$ J mol$^{-1}$K$^{-2}$ 
at $T\simeq0.39\omega_{\rm E}$; 
so that 
it is difficult to reproduce 
a broad hump 
structure of $C/T$ 
observed by experiments.

\begin{figure}
\begin{center}
\epsfxsize=8cm \epsfbox{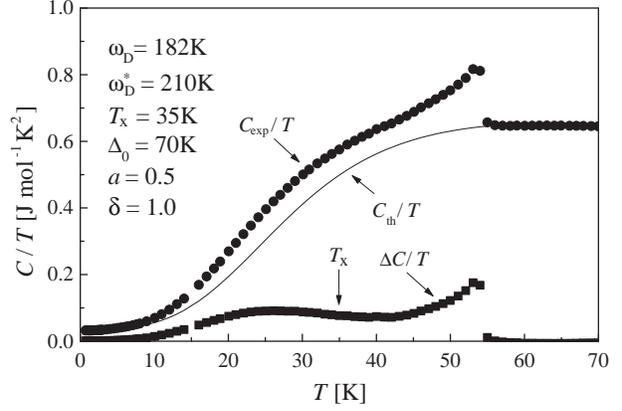} 
\end{center}
\caption{Difference of $C/T$ between experiment and model.
The solid circles represent the experimental values\cite{huxley}, 
solid line theoretical 
ones of model, and the solid squares the differences.  
}
\label{fig:Gamma2}
\end{figure}

\section{Pairing Interaction and Strong-Coupling Effect on $H_{{\rm c}2}$}

%
As we discussed in \S~2, the excess ferromagnetic 
component begins to arise at $T=T_{\rm x}$, where the 
uniform spin susceptibility is diverging due to the mode-coupling effect 
among charge and spin fluctuations with the nesting vector ${\bf Q}$.  
While the charge susceptibility can be also diverging at $T=T_{\rm x}$ 
due to the mode-coupling effect among the uniform charge $N_{{\bf q}=0}$ and 
two modes of SDW, $M_{\bf Q}$ and $M_{-{\bf Q}}$, uniform charge fluctuations 
are suppressed by the long-range Coulomb interaction which works to 
maintain the charge neutrality
as mentioned in \S~2.  
Then, the pairing interaction between quasiparticles with majority spin, 
due to those mode-coupling effects, is mediated mainly by the coupled 
two modes of the spin susceptibility $\chi_{\rm s}$ 
and the charge susceptibility $\chi_{\rm c}$ 
as shown in Fig.~\ref{fig:super}(a), but 
not by the coupled modes of $\chi_{\rm s}$ and $\chi_{\rm s}$. 
The coupled fluctuations of $\chi_{\rm s}$ and $\chi_{\rm c}$ 
which are related to the excess ferromagnetic moment 
are regarded as the 
ferromagnetic 
spin fluctuation $\tilde{\chi}_{\rm s}$ 
which is enhanced for 
the small momentum transfer, i.e., ${\bf q} \sim 0$. 
The important point is that 
the coupled two modes 
can make the scattering wave vectors for the Cooper channel, 
${\bf q}={\bf k}-{\bf k'}$, zero 
which is profitable for the triplet pairing 
(see Fig.~\ref{fig:super}(a)), in marked contrast to the single-mode case
where $\chi_{\rm s}$ or $\chi_{\rm c}$ with finite $\bf Q$ gives 
a small 
repulsive part for the triplet pairing
as has been well recognized~\cite{matsuura,beal-monod,miyake,scalapino}.

These coupled fluctuation propagators give non-zero 
contributions even 
for the external wave vector $q=0$ and 
the bosonic matsubara frequency $\omega_{m}\ne 0$, 
$\omega_{m}\equiv 2m\pi T$.  
If we parameterize these propagators as 
\begin{eqnarray}
\chi_{\rm s}({\bf Q}+{\bf q},{\rm i}\omega_{m})&\simeq&
\chi_{\rm c}({\bf Q}+{\bf q},{\rm i}\omega_{m})
\nonumber \\
&=&\frac{\chi_{0}}{\eta_{0}+A_{0}q^{2}+C_{0}|\omega_{m}|},
\label{barepropagator}
\end{eqnarray}
the propagator $\tilde{\chi}_{\rm s}$ of the coupled fluctuations is given as 
\begin{eqnarray}
\tilde{\chi}_{\rm s}(q,{\rm i}\omega_{m})
& &=T\sum_{\omega_{m'}}\sum_{{\bf q}'}
\chi_{\rm s}({\bf Q}+{\bf q}',\omega_{m'}+\omega_{m})
\nonumber \\
& &\times
\chi_{\rm s}({\bf Q}+{\bf q}',\omega_{m'})
\nonumber
\\
& &\simeq\chi_{0}^{2}
\sum_{{\bf q}'}\int_{0}^{\infty}\frac{dx}{\pi}\coth \frac{x}{2T}
\nonumber \\
& &
\times
\frac{x}{\varphi_{q'}^{2}+x^{2}}
\frac{(\varphi_{{\bf q}+{\bf q}'}+C_{0}|\omega_{m}|)}
{(\varphi_{{\bf q}+{\bf q}'}+C_{0}|\omega_{m}|)^{2}+x^{2}},
\label{CSpropagator}
\end{eqnarray}
where $\varphi_{q}\equiv\eta_{0}+A_{0}q^{2}$.  
It is easy to verify that 
$\lim_{q\to 0}\tilde{\chi}_{\rm s}(q,{\rm i}\omega_{m})$ is finite (non-zero) 
and continuous as a function of $\omega_{m}$ and diverges 
in proportion to 
$T\chi_{0}^{2}(4\pi A_{0})^{-1}/\eta_{0}$ in the 2D system, 
and $T\chi_{0}^{2}(8\pi A_{0}^{3/2})^{-1}/\sqrt{\eta_{0}}$ 
in the 3D system, 
as approaching the onset of coupled CDW-SDW ordering at finite 
temperature, i.e., $\omega_{m}=0$ and $\eta_{0}\to 0$.  
Near the quantum critical point at $T=0$, 
these divergent behaviors are given as 
$\ln(1/\eta_{0})$ in the 2D system, 
and $({\rm const.}-\sqrt{\eta_{0}})$ in the 3D system, respectively.  
It is noted that the latter expression in the 3D system exhibits a sharp 
peak with cusp at $\eta_{0}$, while it does not diverge there.  

By careful inspection of the expression of right hand side of 
eq.~(\ref{CSpropagator}), one can see that the main dependence of 
$\tilde{\chi}_{\rm s}(q,{\rm i}\omega_{m})$ 
on $q$ and $\omega_{m}$ arises through a combination 
$A_{0}q^{2}+C_{0}|\omega_{m}|$.  
An explicit dependence of $\tilde{\chi}_{\rm s}(q,{\rm i}\omega_{m})$ on 
$A_{0}q^{2}+C_{0}|\omega_{m}|$ is not simple compared to those for 
$\chi_{\rm s}$ and $\chi_{\rm c}$, eq.~(\ref{barepropagator}), 
and depends on the space dimensionality.  Indeed, the main part of 
$\tilde{\chi}_{\rm s}(q,{\rm i}\omega_{m})$ is proportional to 
$T\chi_{0}^{2}(2\pi A_{0})^{-1}(A_{0}q^{2}+C_{0}|\omega_{m}|)^{-1}
\ln[(2\eta_{0}+A_{0}q^{2}+C_{0}|\omega_{m}|)/2\eta_{0}]$ in the 2D system, 
and 
$T\chi_{0}^{2}(4\pi A_{0}^{3/2})^{-1}/[\sqrt{\eta_{0}}+
\sqrt{\eta_{0}+(A_{0}q^{2}+C_{0}|\omega_{m}|)/2}] $
in the 3D system.  
The corresponding expressions at $T=0$ near the quantum critical 
point are too complicated to be written here.  

Since the purpose of the present paper is to give a theoretical basis for 
qualitative but unified understanding of the superconducting 
mechanism of UGe$_2$, especially an origin of anomalous temperature 
dependence of $H_{{\rm c}2}(T)$, we adopt the following 
phenomenological form for $\tilde{\chi}_{\rm s}(q,{\rm i}\omega_{m})$ as 
\begin{equation}
\tilde{\chi}_{\rm s}({\bf q},{\rm i}\omega_m)
={\chi_{0}\over \eta+Aq^{2}+C|\omega_{m}|}.  
\label{eliash2}
\end{equation}
This reproduces approximately a characteristic behavior of 
$\tilde{\chi}_{\rm s}(q,{\rm i}\omega_{m})$ near the critical point of 
coupled CDW-SDW ordering if the parameters $\eta$, $A$, and $C$ are 
constants different from $\eta_{0}$, $A_{0}$ and $C_{0}$ in 
eq.~(\ref{barepropagator}) 
and determined so as to reproduce key experimental observations.  

The $q$-dependence of the vertex connecting the coupled CDW-SDW fluctuations 
and the quasiparticles depends on a dynamical structure of the fluctuations.  
If the magnetization is the conserved quantity, the vertex for 
$q=0$ and $\omega_{m}\ne 0$ should vanish as can be verified by a 
perturbational calculation of the triangle diagram 
shown in Fig.~\ref{fig:super}(a).  
However, if the magnetization is not conserved due to the strong 
spin-orbit interaction at U-sites as in the present case, the 
vertex does not vanish even for $q=0$ and $\omega_{m}\ne 0$.  The mode 
mediating the pairing interaction shown in Fig.~\ref{fig:super}(a) 
is that arising from an incoherent process in the sense of the Fermi liquid 
theory.  It is known that such a contribution vanishes in general if 
the relevant quantity of the fluctuations is conserved one~\cite{Leggett}.  
On the other hand, if it is not conserved one, there remains an 
incoherent contribution which arises from high energy processes and 
cannot be described by the response function of quasiparticles.  
It is quite crucial for the behavior of incoherent processes 
whether the relevant quantity is conserved or not, while it is not 
so severe for the quasiparticle response which mainly reflects 
the structure of the phase space of low energy particle-hole continuum 
of quasiparticles as far as the effect of the impurity scattering is 
neglected.

On the basis of the reasoning above, 
the pairing interaction $V({\bf q},{\rm i}\omega_m)$ for 
${\bf q}={\bf k}-{\bf k'}\sim0$ is postulated as 
\begin{equation}
V({\bf q},{\rm i}\omega_m)\simeq U-|I^{*}|^{2}
\tilde{\chi}_{\rm s}({\bf q},{\rm i}\omega_m),
\label{eliash1}
\end{equation}
where $U$ is the short-range ($q$-independent) repulsive interaction, 
$I^{*}$ is the coupling between quasiparticles 
and the coupled CDW-SDW fluctuations.  
The wavenumber dependence 
of $I$ is assumed to be neglected around $q\sim 0$.  
%
The pairing interaction arising from the exchange of 
the single mode of SDW and CDW fluctuations 
with the nesting vector ${\bf Q}$ gives a small repulsive part for 
the triplet p-wave interaction as mentioned above.  
However, this contribution is not 
considered to play a crucial role because the wavenumber dependence of 
$\chi_{\rm s}({\bf q})$ and $\chi_{\rm c}({\bf q})$ is weak at 
${\bf q}\sim 0$.  
So, we neglect its contribution for the moment.  

\begin{figure}
\begin{center}
\epsfxsize=8cm \epsfbox{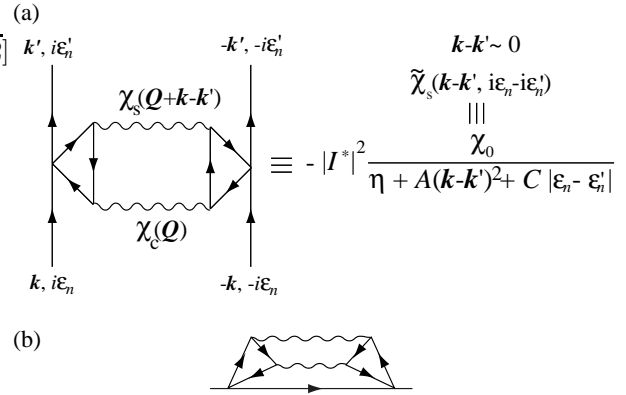} 
\end{center}
\caption{
(a) Paring mechanism mediated by coupled CDW $(\chi_{\rm c})$ and 
SDW $(\chi_{\rm s})$ fluctuations with the nesting vector $\bf Q$. 
(b) Selfenergy of electrons due to one-fluctuation exchange process.
}
\label{fig:super}
\end{figure}

For simplicity of calculations, we assume a three-dimensional 
spherical band with majority spins. 
Then the triplet p-wave component $V(\ell=1)$ of 
$V({\bf q},{\rm i}\omega_m)$ for 
${\bf q}\equiv{\bf k}-{\bf k}^{\prime}$ with 
$|{\bf k}|=|{\bf k}^{\prime}|=k_{\rm F}$ 
is given by 
\begin{equation}
V(\ell=1,{\rm i}\omega_m)\approx -|I^{*}|^{2}
\int_{0}^{\pi}{{\rm d}\theta\over 2}\sin\theta\sqrt{3}\cos\theta
\tilde{\chi}_{\rm s}(q,{\rm i}\omega_m),
\label{eliash3}
\end{equation}
where $q=2k_{\rm F}|\sin\theta/2|$, with $\cos\theta=({\bf k}\cdot
{\bf k}^{\prime})/k_{\rm F}^{2}$.  
Substituting eq.~(\ref{eliash2}) for $\tilde{\chi}_{\rm s}$ and 
after simple algebra, 
we obtain
\begin{eqnarray}
V(\ell=1,{\rm i}\omega_m)&\approx&
-{\sqrt{3}|I^{*}|^{2}\chi_{0}\over 2k_{\rm F}^{2}A}
\nonumber \\ 
&\times&
\biggl[-1+{B+1\over2}\ln\biggl|{(B+2)\over B}\biggr|\biggr],
\label{eliash4}
\end{eqnarray}
where $\omega$-dependent $B$ is defined as 
\begin{equation}
B(\omega_{m})\equiv{\eta+C|\omega_{m}|\over2k_{\rm F}^{2}A}. 
\label{eliash5}
\end{equation}

The corresponding selfenergy $\Sigma$ of electrons due to the 
one-fluctuation exchange process (see Fig.~\ref{fig:super}(b)) is given 
as follows: 
\begin{equation}
\Sigma({\bf p},{\rm i}\epsilon_{n})=T\sum_{\epsilon_{n^{\prime}}}
\sum_{{\bf p}^{\prime}}|I^{*}|^{2}
{\tilde{\chi}_{\rm s}({\bf p}-{\bf p}^{\prime}, 
{\rm i}\epsilon_{n}-{\rm i}\epsilon_{n^{\prime}})\over
{\rm i}\epsilon_{n^{\prime}}-\xi_{{\bf p}^{\prime}}},
\label{eliash6a}
\end{equation}
where $\epsilon_{n}\equiv(2n+1)\pi T$ is the fermionic Matsubara 
frequency.  
The summation in eq.~(\ref{eliash6a}) with respect to ${\bf p}^{\prime}$ 
can be separated approximately and reduced to 
\begin{eqnarray}
\Sigma({\bf p},{\rm i}\epsilon_{n})&=&T\sum_{\epsilon_{n^{\prime}}}
|I^{*}|^{2}\sum_{{\bf p}^{\prime}}{1\over
{\rm i}\epsilon_{n^{\prime}}-\xi_{{\bf p}^{\prime}}}
\nonumber \\
& & \quad \times
\int_{0}^{\pi}{{\rm d}\theta\over 2}\sin\theta
\tilde{\chi}_{\rm s}(q,{\rm i}\epsilon_{n}-{\rm i}\epsilon_{n^{\prime}}), 
\label{eliash6b}
\end{eqnarray}
where $q=2k_{\rm F}|\sin\theta/2|$.  
Here we have approximated as $|{\bf p}|\simeq|{\bf p}^{\prime}|
\simeq k_{\rm F}$ in $\tilde{\chi}_{\rm s}$ in eq.~(\ref{eliash6b}).  
Then, the renormalized frequency 
${\tilde \epsilon}_{n}\equiv \epsilon_{n}+{\rm i}\Sigma({\rm i}\epsilon_{n})$  
is given approximately as follows:  
\begin{eqnarray}
{\tilde \epsilon}_{n}=\epsilon_{n}&+&{\pi}T\sum_{\epsilon_{n^{\prime}}}
{|I^{*}|^{2}\chi_{0}\over2k_{\rm F}^{2}A}
{N_{\rm F}{\rm sign}(\epsilon_{n^{\prime}})\over2}
\nonumber \\
& &\qquad\qquad\times
\ln\biggl|{B(\epsilon_{n}-\epsilon_{n^{\prime}})+2
\over B(\epsilon_{n}-\epsilon_{n^{\prime}})}\biggr|,
\label{eliash6c}
\end{eqnarray}
where $N_{\rm F}$ is the density of states of band electrons.  In deriving 
eq.~(\ref{eliash6c}), we have discarded the real part of 
$\Sigma({\rm i}\epsilon_{n})$ on the understanding that it can be absorbed 
into the chemical-potential shift.  

Here we introduce the function for renormalization effect as 
\begin{equation}
M(\omega_{m})=\lambda{\displaystyle{\ln\biggl|{\Gamma+|\omega_{m}|+2E^{*}
\over \Gamma+|\omega_{m}|}\biggr|}\over
\displaystyle{\ln\biggl|{\Gamma+2E^{*}\over \Gamma}\biggr|}},
\label{eliash7}
\end{equation}
where $\lambda$, $\Gamma$, and $E^{*}$ are defined as 
\begin{equation}
\lambda\equiv
{|I^{*}|^{2}\chi_{0}\over 2k_{\rm F}^{2}A}{N_{\rm F}\over2}
\ln\biggl|{\Gamma+2E^{*}\over \Gamma}\biggr|,
\label{eliash8}
\end{equation}
\begin{equation}
\Gamma\equiv{\eta\over C},  
\label{eliash9}
\end{equation}
and 
\begin{equation}
E^{*}\equiv{2k_{\rm F}^{2}A\over C}. 
\label{eliash9a}
\end{equation}
Here $\Gamma$ is the damping rate of coupled CDW and SDW fluctuation and 
$E^{*}$ is the order of the unrenormalized Fermi energy of band electrons.  

Then, eq.~(\ref{eliash6c}) is written as 
\begin{equation}
{\tilde \epsilon}_{n}=\epsilon_{n}+{\pi}T\sum_{m=-\infty}^{\infty}
M(\epsilon_{n}-\epsilon_{m}){\rm sign}(\epsilon_{m}).
\label{eliash10a}
\end{equation}
After some algebra, we can show that the left hand side of 
eq.~(\ref{eliash10a}) 
is reduced, in the limit of $T\to 0$ and up to the linear term of 
$\epsilon_{n}$, to 
\begin{equation}
{\tilde \epsilon}_{n}=\epsilon_{n}(1+\lambda).  
\label{eliash10}
\end{equation}
Namely, $\lambda$ is the coupling constant giving the mass enhancement factor.  
Of course, there may exist other contributions to the mass enhancement 
arising from single-mode processes of exchanging 
spin and charge fluctuations with large nesting vector ${\bf Q}$.  
However, this contribution is expected to give only moderate 
enhancement for the Sommerfeld constant if the fluctuations are dominated 
by the nesting as has been discussed elsewhere\cite{maebashi}.  
So, we neglect this contribution for simplicity of analysis.

In terms of $\lambda$, $\Gamma$ and $E^*$, the pairing interaction of p-wave 
($\ell$=1) is written as 
\begin{eqnarray}
& &V(\ell=1,{\rm i}\omega_m)\approx 
-{\sqrt{3}\over N_{\rm F}}{2\lambda\over \displaystyle
{\ln\biggl|{\Gamma+2E^{*}\over \Gamma}\biggr|}}
\nonumber \\
& &\times
\biggl[
-1+{E^{*}+\Gamma+|\omega_{m}|\over2E^{*}}
\ln\biggl|{2E^{*}+\Gamma+|\omega_{m}|\over
\Gamma+|\omega_{m}|}\biggr|\biggr].
\label{eliash10b}
\end{eqnarray}

On the derivation above, the pairing interaction $V$ for a relevant channel 
can be expressed in a separable form 
in the wave vector and frequency space as 
\begin{equation}
V({\bf k}-{\bf k}^{\prime},{\rm i}\epsilon_{n}
-{\rm i}\epsilon_{n^{\prime}})
=\phi({\bf k})\phi({\bf k}^{\prime})
v({\rm i}\epsilon_{n}-{\rm i}\epsilon_{n^{\prime}}),
\label{Eliash11}
\end{equation}
where $\phi({\bf k})$ is the wavefunction of relevant pairing 
channel, and $v$ represents the frequency dependence of interaction 
due to the retardation.  In the present case, 
$\phi({\bf k})\simeq \sqrt{3}{\hat k}_{z}$, and $v$ is given by 
eq.~(\ref{eliash10b}).  
Thus, the gap function $\Delta$ also has the separable form 
$\Delta({\bf k},{\rm i}\epsilon_{n})
=\phi({\bf k})\Delta({\rm i}\epsilon_{n})$ 
and satisfies the gap equation 
\begin{eqnarray}
\Delta({\rm i}\epsilon_{n})&=&-T\sum_{n^{\prime}}
\sum_{{\bf k}^{\prime}}
v({\rm i}\epsilon_{n}-{\rm i}\epsilon_{n^{\prime}})
\nonumber \\
& &\quad\times
{|\phi({\bf k}^{\prime})|^{2}\Delta({\rm i}\epsilon_{n^{\prime}})
\over \xi^{2}_{{\bf k}^{\prime}}+{\tilde{\epsilon}}_{n^{\prime}}^{2}
+|\phi({\bf k}^{\prime})\Delta({\rm i}\epsilon_{n^{\prime}})|^{2}},
\label{Eliash12}
\end{eqnarray}
where $\xi_{{\bf k}}$ is the dispersion of quasiparticles without the 
renormalization due to the many-body effect of eq.~(\ref{eliash6b}), and 
${\tilde{\epsilon}}_{n}$ is 
the renormalized frequency due to the many-body effect of eq.~(\ref{eliash6c}).  
Explicit form of $\tilde{\epsilon}_{n}$ in the present model is given by 
eq.~(\ref{eliash10a}).  

The linearized gap equation is then given as follows:
\begin{equation}
\Delta({\rm i}\epsilon_{n})=-T\sum_{n^{\prime}}
v({\rm i}\epsilon_{n}-{\rm i}\epsilon_{n^{\prime}})
\Delta({\rm i}\epsilon_{n^{\prime}})
\sum_{{\bf k}^{\prime}}
{|\phi({\bf k}^{\prime})|^{2}
\over \xi^{2}_{{\bf k}^{\prime}}+{\tilde{\epsilon}}_{n^{\prime}}^{2}}.
\label{Eliash13}
\end{equation}
At this stage, it may not be so bad to approximate $|\phi({\bf k})|^{2}$ 
as an average $\langle|\phi({\bf k})|^{2}\rangle$ near the Fermi level.  
This quantity 
is nearly equal to unity due to the normalization condition.  
Namely, eq.~(\ref{Eliash13}) can be approximated by the same one for 
the conventional s-wave pairing.  

%
So far, we have constructed the general framework of the pairing mechanism 
mediated by coupled CDW and SDW fluctuations. 
To discuss the real materials rigorously, it is necessary to take into 
account of the band structure and the Fermi surface, accurately. 
Hereafter, we would like to 
discuss the anomalous behaviors of $H_{{\rm c}2}(T)$ under 
pressures in $\rm UGe_2$. 
However, there is 
not enough
information about the band structure and the Fermi surface 
responsible for the superconductivity 
under pressure and magnetic field in $\rm UGe_2$ at present. 
Hence, we proceed the discussion on the basis of the 
above framework with three-dimensional spherical band 
and consider the features of $\rm UGe_2$ 
by taking into account of the most-important factors such as 
pressure and magnetic-field dependences of 
mass-enhancement factor and $T_{\rm x}$ 
as will be discussed below. 
It is an important task to consider the accurate band structure 
and to determine the pairing symmetry of the orbital part
in the next stage of the theoretical approach. 
However, we think that 
the anomalous behaviors of $H_{{\rm c}2}$ under pressures 
do not depend on such details 
but are caused by the key factors which will be explained below.
%

Let us summarize again the behaviors of upper-critical field under 
pressures (see Fig.~\ref{fig:Hc2}) and give its intuitive understanding.  
At $P=11.4$ kbar, $H_{{\rm c}2}$ shows only gradual increase 
in spite that $T_{\rm SC}$ was originally high at $H=0$. 
This is because $T_{\rm x}$, around which the attractive interaction is 
induced, 
goes up and the enhanced fluctuations associated with $T_{\rm x}$ fades away 
as $H$ increases.  
On the other hand, at $P=13.5$ kbar $T_{\rm SC}$ is less than the maximum 
value at $H=0$.  
However, as $H$ increases, $T_{\rm x}$ goes up to be close 
and the attractive interaction is enhanced so that $H_{\rm c2}$ grows rapidly. 
Then, $H_{\rm c2}$ at $P=13.5$ kbar crosses above that of $P=11.4$ kbar.
As for $P=15.3$ kbar it is interpreted that 
$H_{\rm c2}$ reaches $T=0$ before $T_{\rm x}$ closes up 
under magnetic field 
and thereby shows only the gradual increase. 

The strong-coupling formalism for $H_{{\rm c}2}$ originally developed 
for s-wave superconductors on the electron-phonon 
mechanism~\cite{helfand,schossmann}, can be extended rather easily 
for the present case.  
It is because the effect of magnetic field can be 
incorporated into the theory just as in the conventional case as far as 
the vector potential only couples with the center of mass coordinate of 
the wave function of the Cooper pair, but does not affect its relative 
coordinate.  This is the case in usual situations since the size of the 
Cooper pair is far smaller than the width of the Landau orbital except for 
$H\simeq H_{{\rm c}2}(T=0)$.  Thus we can use the strong-coupling formalism 
for $H_{{\rm c}2}$ developed for treating the s-wave pairing on a slight 
modification.  

Starting from the Eliashberg equations, eq.~(\ref{eliash10a}) and 
eq.~(\ref{Eliash12}), one obtains 
the linearized gap equation under the magnetic field $H$ as follows: 
\begin{eqnarray}
\tilde{\Delta}(i\epsilon_{n})= \frac{T}{N_{\rm F}}
\sum_{m}\tilde{V}(\epsilon_{n}-\epsilon_{m})K(\tilde{\epsilon}_{m})
\tilde{\Delta}(i\epsilon_{m}), 
\label{eq:gap}
\end{eqnarray}
where the effective coupling $\tilde{V}$ is given by 
\begin{equation}
\tilde{V}(\omega_{m}) = -N_{{\rm F}}V(\ell=1,{\rm i}\omega_{m}),
\label{eq:lamda}
\end{equation}
where $V(\ell=1,{\rm i}\omega_{m})$ is defined by eq.~(\ref{eliash10b}).  
The integration kernel $K$ in eq.~(\ref{eq:gap}) is given as 
\begin{eqnarray}
K(\tilde{\epsilon}_{n})&=&\frac{4{\pi}N_{\rm F}}{\sqrt{2e{v_{\rm F}}^2H}}
\int_{0}^{\infty} dq {\exp}(-q^{2})
\nonumber \\
& & \times \tan^{-1} \left(
\frac{q \sqrt{2e{v_{\rm F}}^2H}}
{2|\tilde{\epsilon}_{n}|}
\right),
\label{eq:chi}
\end{eqnarray} 
where $e$ is the elementary charge, 
$v_{\rm F}$ is the Fermi velocity of band electrons and 
$\tilde{\epsilon}_{n}$ is defined by eq.~(\ref{eliash10a}).  
In the present case where only the electrons with majority spin 
participate in the formation of the SC state, the non-unitary 
triplet pairing is 
expected. Then the paramagnetic effect due to the Zeeman splitting 
has been neglected in deriving eq.~(\ref{eq:chi}).  
The strong-coupling effect is included in the frequency dependence 
of the gap function $\tilde{\Delta}({\rm i}\epsilon_{n})$ in 
eq.~(\ref{eq:gap}) and ${\tilde \epsilon}_n$ in eq.~(\ref{eq:chi}).  
The upper critical field $H_{{\rm c}2}$ is determined by the condition that 
the linearized gap equation starts to have a nontrivial solution 
$\tilde{\Delta}({\rm i}\epsilon_{n}) \ne 0$ as $H$ is decreased with $T$ fixed.
The resultant equation to be solved is 
\begin{eqnarray}
\det \left[
\delta_{nm}- \frac{T}{N_{\rm F}}\tilde{V}(\epsilon_{n}-\epsilon_{m}) 
K(\tilde{\epsilon}_{m})\right]=0, 
\label{eq:det}
\end{eqnarray}
where $\delta_{nm}$ is the Kronecker delta symbol.  

Our strategy for discussing $H_{{\rm c}2}(T)$ is as follows:  
First, we determine the pressure dependence of $\lambda$ in 
eq.~(\ref{eliash7}), 
giving the mass enhancement factor, by experimental 
data of specific heat\cite{tateiwa}.  
Second, we estimate the pressure 
dependence of parameters $\Gamma$, eq.~(\ref{eliash9}), 
and $E^{*}$, eq.~(\ref{eliash9a}), so as 
to reproduce $T_{\rm SC}$.  
Third, we deduce the magnetic-field dependence of 
the parameters with the use of a functional form of $T_{\rm x}(P,H)$ experimentally 
obtained\cite{tateiwa2,sheikinprivate}.  Then, $H_{{\rm c}2}(T)$ is calculated 
from eq.~(\ref{eq:det}).  

First of all, we discuss the $P$-dependence of $\lambda$ at $H=0$.  
In Fig.~\ref{fig:gmfit} open circles with an error bar denote the ratio 
of the specific-heat coefficient under finite and zero pressures, 
$\gamma(P)/\gamma(0)$, 
which are measured at $H=0$~\cite{tateiwa}.  
This ratio can be written in terms of the 
mass-enhancement factor as 
$\gamma(P)/\gamma(0)=[1+{\lambda}(P)]/[1+{\lambda}(0)]$.  
The enhancement of $\gamma(P)$ for $P \gsim 13$ kbar 
is considered to be contributed by two factors: 
one is the fluctuation associated with the anomaly at $T_{\rm x}$ and 
the other is the critical FM fluctuation near $T_{\rm C}$ 
which we do not consider here as irrelevant for the superconductivity.  
We consider the former is essential for inducing the SC state.  
Hence, we introduce a biquadratic function 
of $[1+{\lambda}(P)]/[1+{\lambda}(0)]$ around $P=P_{\rm c}\simeq 13$ kbar 
to fit $\gamma(P)/\gamma(0)$ 
as the dashed line in Fig.~\ref{fig:gmfit}, 
although experimental data for $\gamma(P)$ remains to increase gradually 
even in the $P \gsim 13$ kbar region.  
We assume that the mass enhancement at zero pressure gives no contribution to 
the SC state for $10 \lsim P \lsim 17$ kbar, i.e., we set $\lambda(0)=0$. 
In this way, we now obtain the pressure dependence of the mass-enhancement 
factor, $\lambda(P)$ in eq.~(\ref{eliash7}), which is denoted by the shaded 
line in Fig.~\ref{fig:gmfit}.

\begin{figure}
\begin{center}
\epsfxsize=8cm \epsfbox{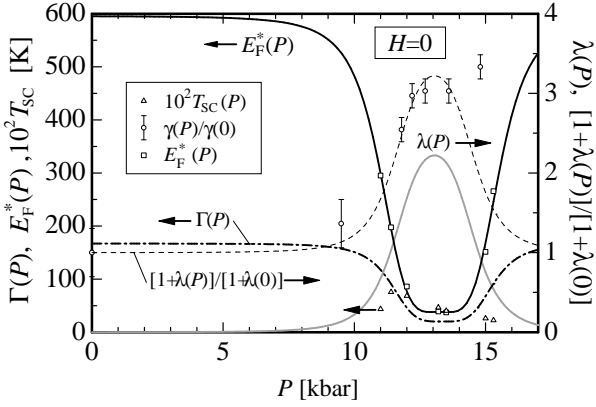}
\end{center}
\caption{
Pressure dependence of parameters at $H=0$.
Open circles with an error bar denote the ratio $\gamma(P)/\gamma(0)$ 
measured without magnetic field\cite{tateiwa}. 
Open triangles denote ten times of 
the SC transition temperature $T_{\rm SC}$\cite{huxley}.
Open squares denote $E^{*}(P)$ which reproduce $T_{\rm SC}$ by using 
$\lambda(P)$ and $\Gamma(P)$.  
The dashed line and the shaded line denote 
$[1+\lambda(P)]/[1+\lambda(0)]$ and $\lambda(P)$, respectively.
The dash-dot line and the solid line denote $\Gamma(P)$ and $E^{*}$, 
respectively. 
}
\label{fig:gmfit} 
\end{figure}

Secondly, we determine the $P$ dependence of $\Gamma$, eq.~(\ref{eliash9}), 
and $E^{*}$, eq.~(\ref{eliash9a}), so as to reproduce $T_{\rm SC}(P)$ at $H=0$ 
by using $\lambda(P)$ determined as above.  
In Fig.~\ref{fig:gmfit} open triangles denote 100 times of 
the SC transition temperature, i.e., $10^{2}T_{\rm SC}$.  
We assume the shape of $\Gamma(P)$ as the dash-dotted line in Fig.~\ref{fig:gmfit}, 
which becomes narrow toward the critical pressure $P\sim13.5$ kbar. 
The open squares denote the values of $E^*$ 
which reproduce the experimental data of $T_{\rm SC}$ at each pressure. 
Here, we introduce a biquadratic function of $E^{*}(P)$ 
as the solid line in Fig.~\ref{fig:gmfit} 
to fit discrete data denoted by open squares.  
Then we obtain the $P$-dependence of $\lambda$, eq.~(\ref{eliash8}), 
$\Gamma$, eq.~(\ref{eliash9}), and $E^{*}$, eq.~(\ref{eliash9a}) 
from the experimental data of 
$\gamma$ and $T_{\rm SC}$ in the $H=0$ case.  
If we choose another shape of $\Gamma(P)$, the shape of $E^{*}(P)$ 
also changes to reproduce $T_{\rm SC}$. 
However, the spectrum of coupled CDW and SDW fluctuations which mediate 
the superconductivity is considered to be sharp at $T_{\rm x}$ 
and thereby we here adopt the function $\Gamma(P)$ as shown in Fig.~\ref{fig:gmfit}. 

Thirdly, we discuss the $H$-dependence of the parameters $\lambda$, $\Gamma$, 
and $E^{*}$.  Here we note the $H$-dependence of $T_{\rm x}$ is 
identified  by the rapid change of the slope of the resistivity 
$\rho(T)$ under the magnetic 
field at each pressure~\cite{Jflouquet,tateiwa2} 
and is shown in Fig.~\ref{fig:TxH}(a)~\cite{Jflouquet}.  
We see that $T_{\rm x}$ depends on $H$ linearly at $P=11.4$ kbar and 
increases steeply at $P=13.5$ kbar.  
Incorporating these behaviors, we introduce a function $T_{\rm x}(P,H)$ 
as shown in Fig.~\ref{fig:TxH}(b). 
The key point is that $T_{\rm x}$ is an increasing function of $H$ 
and a decreasing function of $P$.  
By using this form of function $T_{\rm x}(P,H)$, 
the $H$ dependence of $\lambda$, eq.~(\ref{eliash8}), 
and $\Gamma$, eq.~(\ref{eliash9}), and $E^{*}$, eq.~(\ref{eliash9a}), 
is taken into account in the following way: 
We can obtain the pressure $\tilde{P}$ at $H=0$ 
which gives the same value of $T_{\rm x}$ at $P$ and $H$ 
by using the relation 
\begin{eqnarray}
T_{\rm x}(\tilde{P},H=0)=T_{\rm x}(P,H). 
\label{eq:TxPH}
\end{eqnarray}
This always gives $\tilde{P} \le P$ for $H \ge 0$.  
Namely, 
applying the positive magnetic-field turns out to be equivalent to 
the negative pressure at $H=0$.  
In this way we can take into account the effect of the magnetic field 
on parameters 
$\lambda$, $\Gamma$ and $E^{*}$ which appear in 
the gap equation eq.~(\ref{eq:det}) 
of our model.

\begin{figure}
\begin{center}
\epsfxsize=8cm \epsfbox{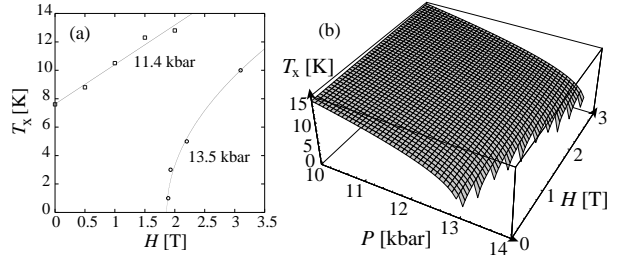} 
\end{center}
\caption{(a) Experimental data of $T_{\rm x}$ 
as a function of filed applied parallel to the crystallographic a-direction 
at $P=11.4$ kbar and $P=13.5$ kbar~\cite{Jflouquet}. 
(b) Introduced function of $T_{\rm x}(P,H)$ to calculate $H_{\rm c2}$ at each 
pressure. 
}
\label{fig:TxH}
\end{figure}

On these estimation of parameters, we can calculate $H_{\rm c2}$ 
at each pressure as follows: 
The pressure $P$ and the magnetic field $H$ are fixed. 
With use of eq.~(\ref{eq:TxPH}), the reduced pressure 
$P=\tilde{P}$ is determined.  Then, the linearized gap equation 
eq.~(\ref{eq:det}), with parameters $\lambda(\tilde{P})$, $\Gamma(\tilde{P})$ 
and $E^{*}(\tilde{P})$ introduced in Fig.~\ref{fig:gmfit}, is solved to obtain 
$H_{\rm c2}$.  The Fermi energy $\epsilon_{\rm F}$ and 
the Fermi velocity $v_{\rm F}$ of unrenormalized band electrons have been 
set to reproduce the order of magnitude of observed $H_{{\rm c}2}(T\to 0)$.  

Results of $H_{\rm c2}(T)$ for $P=11.4$, 
13.5 and 15.3 kbars are shown in Fig.~\ref{fig:rslt}.  
At $P=11.4$ kbar, where $T_{\rm SC}(H=0)$ is nearly maximum, 
$H_{\rm c2}(T)$ merely increases gradually and monotonously as $T$ decreases.  
This is because applying the magnetic field gives negative pressure 
which makes $T_{\rm SC}(H)$ decreased, as seen in Fig.~\ref{fig:gmfit}.  
This reflects the fact that $T_{\rm x}(H)$ goes up and away from 
$T_{\rm SC}(H=0)$ as $H$ increases at $P=11.4$ kbar.  
On the other hand, at $P=13.5$ kbar, where $T_{\rm SC}(H=0)$ is rather 
smaller than the maximum at $P=11.4$ kbar, $H_{\rm c2}(T)$ increases 
steeply and shows first-order like ``transition" at $H=1.8$ T.  
This anomalous increase is understood  
as a result of sudden appearance of $T_{\rm x}$ at $H=1.8$ T 
due to the magnetic field as shown in Fig.~\ref{fig:TxH}(a).  
Alternatively, such an increase can be understood by the fact that 
the applied magnetic field has an effect of negative pressure which 
makes $T_{\rm SC}(H=0)$ shift to the maximum value at $P=11.4$ kbar 
from $P=13.5$ kbar as shown in Fig.~\ref{fig:gmfit}.  
From these results we can see that the crossing of $H_{\rm c2}(T)$ at 
$P=11.4$ kbar and $P=13.5$ kbar is caused by two counter effects of 
the magnetic field: The suppression of $H_{\rm c2}(H)$ at $P=11.4$ kbar 
and the enhancement of $H_{\rm c2}(H)$ at $P=13.5$ kbar.  
At $P=15.3$ kbar $H_{\rm c2}$ shows the gradual and monotonous increase 
as T decreases.  It is because the magnetic field $H\le H_{{\rm c}2}(T=0)$ 
is not enough to recover $T_{\rm x}(H)>0$ so that the concave $T$-dependence 
of $H_{{\rm c}2}(T)$ is not realized.  

If the $H$ dependence of $T_{\rm x}$ at $P=13.5$ kbar is gradual 
rather than the square-root-like behavior used in the present calculation 
as in Fig.~\ref{fig:TxH}(b) 
(for example, if $T_{\rm x}$ has a linear dependence with respect to $H$), 
the first-order-like jump in $H_{\rm c2}$ does not appear. 
However, $H_{\rm c2}(T)$ 
still increases steeply and crosses that of $P=11.4$~kbar. 
Then, the detail of the shape of $H_{\rm c2}(T)$ changes according to the 
$H$ dependence of $T_{\rm x}$, but the global feature of $H_{\rm c2}(T)$ under 
pressure such as crossing does not change.

\begin{figure}
\begin{center}
\epsfxsize=8cm \epsfbox{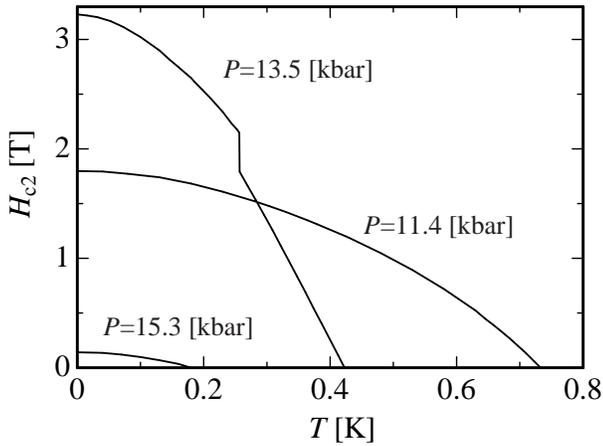} 
\end{center}
\caption{
The upper critical field $H_{\rm c2}$ calculated based on the 
strong-coupling-superconductivity formalism explained in the text 
at $P=11.4$ kbar, $13.5$ kbar and $15.3$ kbar. 
}
\label{fig:rslt}
\end{figure}

The above result of $H_{{\rm c}2}(T)$ is qualitatively the same as that 
obtained on a more simplified phenomenological model in which the effective 
coupling $\lambda(\omega_{m})$, eq.~(\ref{eliash8}), and the function for 
renormalization effect $M(\omega_{m})$, eq.~(\ref{eliash7}), are approximated as 
\begin{equation}
\lambda(\omega_{m})\approx M(\omega_{m})
\approx{\lambda \over N_{\rm F}}{\Gamma\over \Gamma+|\omega_{m}|}. 
\label{lorentzian}
\end{equation}
As shown in ref~\cite{Jflouquet,sheikin}, this model explains the crossing of 
$H_{{\rm c}2}(T)$ at $P=11.4$ and 13.5 kbars when the parameters 
$\lambda$ and $\Gamma$ are fixed in a way similar to the present paper.

\section{Summary}
We have discussed that characteristic aspects of ferromagnetic 
superconductor UGe$_2$ 
can be understood in a unified way on a single assumption that 
the superconductivity is mediated by the 
coupled CDW and SDW fluctuations 
originating from the CDW ordering of the majority-spin band 
at $T=T_{\rm x}$.  
The growth of extra magnetization below $T<T_{\rm x}$ is explained 
by a mode-coupling effect of the coupled CDW and SDW ordering 
caused by the imperfect nesting of the majority-spin band.  
The anomalously large entropy around $T=T_{\rm x}(P=0)$ 
is understood by the Kohn effect of softened optical phonons 
due to the onset of the CDW ordering of majority-spin band.

We have constructed the general framework of the pairing mechanism 
mediated by the coupled CDW and SDW fluctuations. 
The pairing interaction in the ``p"-wave channel among quasiparticles 
in the majority-spin band is shown to induce the {\it ferromagnetic} 
fluctuations enhanced around $P=P_{\rm x}$ due to the coupled 
CDW and SDW fluctuations.  
%
This mechanism also provides a natural explanation for the fact 
that the superconducting phase appears inside the ferromagnetic phase 
in the temperature-pressure ($T$-$P$) phase diagram. 
%
The anomalous temperature dependence 
of the upper critical field $H_{{\rm c}2}(T)$ observed at 
$P=11.4$ and $P=13.5$ kbars 
is understood by the strong-coupling formalism and the experimental 
fact that $T_{\rm x}$ is an increasing function of $H$.  

Of course, there still remains many aspects to be clarified both 
experimentally and theoretically.  A direct measurement of coupled CDW and SDW 
ordering at $T<T_{\rm x}$ is indispensable.  
In particular, the calculation of $\chi({\bf q})$ based on the band-structure 
calculation and the search of CDW ordering by elastic neutron 
and X-ray scattering measurement is desired especially at 
around $P=9$ kbar where 
the decrease of $\rho(T)$ at $T=T_{\rm x}$ is much more prominent compared to 
the ambient pressure.  
Soft modes of phonons toward $T=T_{\rm x}$ should be observed at the wave vector 
$\bf Q$ where the nesting occurs. 
Origin of a first-order like ferromagnetic 
transition at $T=T_{\rm C}$ has not yet been clarified.  The spontaneous 
magnetic moment, in the limit $T\to 0$, also exhibits a first-order 
like transition at $P=P_{\rm x}~$\cite{huxley2}.  This is also to be clarified.  
The identification of the type of the superconducting order parameter is also 
important task left in the future. 
To identify the order parameter the group-theoretical classification 
is usually done.  
However, 
it is necessary to clarify the relevant effective interaction 
which works quasiparticles 
to identify the most stable order parameter 
among the possible candidates~\cite{machida,fomin}. 
%
We suppose that the superconducting gap opens at the parts of the Fermi surface 
with crystalline inversion symmetry, 
which is different from those with the CDW gap to gain both the condensation energies 
of CDW and superconducting states.  
The steep decrease of the superconducting temperature, $T_{\rm SC}$ for 
the $P<P_{\rm x}$ region in the $T$-$P$ phase diagram 
may be due to the partial collapse of the Fermi surface by 
the imperfect nesting. 
%

Quite recently, it has been discovered that ferromagnetic 
metal URhGe, at $T<10$ K, sets in superconducting state 
at $T\simeq0.35$~K under the ambient pressure~\cite{aoki}.  
This compound seems to share common aspects with UGe$_2$, such as 
the crystal structure with zigzag chain of U atoms, existence of a 
small hump of $C/T$ at $T\sim T_{\rm C}/3$ as seen in Fig.~\ref{fig:Gamma2}, 
enhanced ratio $A/\gamma^{2}$ 
of 10 times larger than the so-called Kadowaki-Woods ratio\cite{aoki2}, 
suggesting that 
the system is located near the quantum critical point in a kind of another, 
and of course the coexistence of superconductivity and ferromagnetism.  
The experimental evidence of the CDW has not been 
reported in $\rm URhGe$ at present.
However, we expect that the small hump of $C/T$ at $T \sim T_{\rm C}/3$ 
may be the signature of the CDW transition, 
since the zig-zag-chain structure of U atoms, 
which is a common feature of $\alpha$-Uranium, 
$\rm UGe_{2}$ and $\rm URhGe_{2}$, 
provides the low dimensionality of the Fermi surface which enlarges 
the nesting instability. 
Further studies of URhGe as well as $\rm UGe_{2}$ 
are expected to extend the frontier of a class of 
ferromagnetic superconductor. 

\section*{Acknowledgments}
This work started from conversations with members of Flouquet group at 
DRFMC of CEA/Grenoble when one of us (K.M.) was staying there.  We have 
much benefited from 
discussions with A. Huxley, I. Sheikin, D. Braithwaite, and J. Flouquet.  
Unpublished experimental data they showed us prior to publication was 
essential for us to develop a theoretical idea.  
We acknowledge N. Tateiwa, T. C. Kobayashi, and Y. \=Onuki 
for allowing us to use their experimental data of specific heat as well as 
fruitful discussions.  
We also acknowledge H. Yamagami for providing his result of 
band structure calculations  
prior to publication as well as valuable discussions.  
K.M. has benefited from conversations with G. Lonzarich, V. P. Mineev, and 
C. Pfleiderer on the pairing interaction, which led us to clarify the 
presentations in \S4.  
S.W. thanks M. Sigrist for useful discussion. 
Thanks are also due to 
D. Aoki, 
K.~Hori, G. Oomi, N. K. Sato, R.~Settai and H.~Kohno 
for useful discussions.  
This work was supported by the Grant-in-Aid for COE Research 
(No.10CE2004) from Monbu-Kagaku-sho, and in part by a Grant-in-Aid 
for Scientific Research for Specified Area (No.12046246) by the Japan 
Society for Promotion of Science.



\end{document}